\documentstyle[12pt]{article}

\begin{document}

\title{{\Large {\bf Note on Matter Collineations in Kantowski-Sachs, Bianchi Types
I and III Spacetimes }}}
\author{Pantelis S. Apostolopoulos and Michael Tsamparlis \\
{\small {\it Department of Physics, Section of
Astrophysics-Astronomy-Mechanics},}\\
{\small {\it University of Athens, Panepistemiopolis, Athens 157 83, GREECE}}}
\maketitle

\begin{abstract}
We show that the classification of Kantowski-Sachs, Bianchi Types I and III
spacetimes admitting Matter Collineations (MCs) presented in a recent paper
by Camci et al. [Camci, U., and Sharif, M. {\em Matter Collineations in
Kantowski-Sachs, Bianchi Types I and III Spacetimes}, (2003) {\em Gen.
Relativ. Grav. }{\bf 35} 97-109] is incomplete. Furthermore for these
spacetimes and when the Einstein tensor is non-degenerate, we give the {\em %
complete} Lie Algebra of MCs and the algebraic constraints on the spatial
components of the Einstein tensor.
\end{abstract}


\section*{Introduction}

In a recent paper Camci et al. \cite{Camci-Sharif} studied Matter
Collineations (MCs) in Kantowski-Sachs ($k=+1$), Bianchi Type I ($k=0$) and
Bianchi Type III ($k=-1$) spacetimes, which are described by the following,
non static, hypersurface homogeneous Locally Rotationally Symmetric (LRS)\
metrics: 
\begin{equation}
ds^2=-dt^2+A^2(t)dx^2+B^2(t)\left[ dy^2+\Sigma ^2(y,k)dz^2\right]
\label{sx1.1}
\end{equation}
where $\Sigma (y,k)=\sin y,\sinh y,y$ and $k=1,-1,0$ is the curvature of the
Euclidean 2-space of constant curvature.

The main conclusions of their interesting work are:

\begin{enumerate}
\item  There are degenerate cases of $T_{ab}$ which admit a finite number of
proper MCs and more specifically 9 ($k=0$) or 10 MCs ($k=\pm 1$).

\item  In case where $T_{ab}$ is non-degenerate there are either {\em six}
proper MCs or {\em no }proper MCs.
\end{enumerate}

Unfortunately some of the results of the aforementioned paper are incorrect.
Indeed:

a. The first conclusion is true for $k=\pm 1$ but not for $k=0$.

b. The second conclusion is incomplete, in the sense, that there are cases
which the authors did not consider. Therefore the classification they give
is not complete. Furthermore the results given for the non-degenerate case $%
k=0$ are incorrect.

The purpose of this note is twofold:

A. To show that in the degenerate case and for $k=0$ the dimension of the
Lie Algebra of MCs is {\em 5, }therefore there exists (possibly) only {\em %
one } proper MC.

B. For the non-degenerate case to give the correct and {\em complete} Lie
Algebra of MCs and the differential constraints on the components of the
Einstein tensor.

For later use we give the Einstein tensor $G_{ab}$ for the metric (\ref
{sx1.1}): 
\begin{equation}
G_{00}=G_0=2\frac{\dot{A}\dot{B}}{AB}+\left( \frac{\dot{B}}B\right) ^2+\frac
k{B^2}  \label{sx1.2}
\end{equation}
\begin{equation}
G_{11}=G_1=-A^2\left[ 2\frac{\ddot{B}}B+\left( \frac{\dot{B}}B\right)
^2+\frac k{B^2}\right]  \label{sx1.3}
\end{equation}
\begin{equation}
G_{22}=G_2=-B^2\left[ \frac{\ddot{A}}A+\frac{\ddot{B}}B+\frac{\dot{A}\dot{B}%
}{AB}\right] =\Sigma ^{-2}(y,k)G_{33}  \label{sx1.4}
\end{equation}
where a dot denotes differentiation w.r.t. $t$ and $k=0,1,-1$.

\section*{The degenerate case}

We shall only consider case (c.i) of \cite{Camci-Sharif} for which $G_0=0$.
Equation (\ref{sx1.2}) for $k=0$ implies: 
\begin{equation}
2\frac{\dot{A}\dot{B}}{AB}+\left( \frac{\dot{B}}B\right) ^2=0.  \label{sx1.5}
\end{equation}
Demanding $\dot{B}\neq 0$ (rank$G_{ab}=3$) equation (\ref{sx1.5}) gives: 
\begin{equation}
B=\frac D{A^2}  \label{sx1.6}
\end{equation}
where $D$ is a constant of integration.

Using equation (\ref{sx1.6}) in (\ref{sx1.3}) and (\ref{sx1.4}) we obtain ($%
k=0$): 
\begin{equation}
G_1=4\left( \ddot{A}A-4\dot{A}^2\right)  \label{sx1.7}
\end{equation}
\begin{equation}
G_2=\frac{D^2\left( \ddot{A}A-4\dot{A}^2\right) }{A^6}.  \label{sx1.8}
\end{equation}
As correctly stated in \cite{Camci-Sharif} the vector fields ${\bf \xi }%
_{(5)},{\bf \xi }_{(6)},{\bf \xi }_{(7)},{\bf \xi }_{(8)},{\bf \xi }_{(9)}$
(equation (41) of \cite{Camci-Sharif}) are MCs of the metrics (\ref{sx1.1})
provided the components of $G_{ab}$ satisfy the condition $G_1=\epsilon G_2$
where $\epsilon $ is a constant. Replacing equations (\ref{sx1.7}) and (\ref
{sx1.8}) in this condition we find ($\ddot{A}A-4\dot{A}^2\neq
0\Leftrightarrow G_1\neq 0$): 
\[
4\left( \ddot{A}A-4\dot{A}^2\right) =\epsilon \frac{D^2\left( \ddot{A}A-4%
\dot{A}^2\right) }{A^6} 
\]
which implies that $A=$constant, therefore the spacetime reduces to
Minkowksi spacetime i.e. $G_{ab}=0$.

The error lies in the solution of equations (11) and (12) of \cite
{Camci-Sharif}. Indeed from these two equations we obtain the general
condition: 
\begin{equation}
G_1=A_1G_2^{\alpha _1}  \label{sx1.9}
\end{equation}
where $A_1$ and $\alpha _1$ are constants. Therefore for $\alpha _1\neq 1$
there exists only the vector field: 
\begin{equation}
{\bf X}=-2\frac{G_1}{\dot{G}_1}\partial _t+x\partial _x+\frac y{\alpha
_1}\partial _y  \label{sx1.10}
\end{equation}
which {\em may be proper MC} (what it is will follow from the solution of
the constraint equation (\ref{sx1.9})). Therefore in this case the dimension
of the Lie Algebra of MCs is {\em five} ({\em one proper}) and not nine as
the authors claim. The vector field (\ref{sx1.10}) is given in \cite
{Camci-Sharif} with the contradictory restriction $\alpha _1=1$.

It is to be noted that the physical interest in the degenerate case is
limited, because it is well known that the only interesting case of
degenerate stress-energy tensor is when $rankG_{ab}=1$ in which case the
matter is either dust fluid or radiation and null Einstein-Maxwell fields 
\cite{Carot-daCosta-Vaz}. In this case the $C^\infty $ MCs for the metrics (%
\ref{sx1.1}) form an infinite dimensional Lie Algebra \cite{Hall-Roy-Vaz}.

\section*{The non degenerate case}

In the non-degenerate case, $rankG_{ab}=4$ and $G_{ab}$ can be treated as a
metric \cite{Hall-Roy-Vaz}. This means that the Lie algebra ${\cal C}$ of
MCs is finite dimensional with possible dimension {\em 4,5,6,7,10}. Four of
these vectors are the KVs of the metric (\ref{sx1.1}), therefore there can
be either {\em 0,1,2,3,6} {\em proper} MCs. The authors have obtained the
MCs only for the cases where dim${\cal C}={\em 0,6}$ and have ommited the
rest. However even when dim${\cal C}=10$ there are problems concerning the
forms of the MCs given in \cite{Camci-Sharif}. In order to justify our
claims we present the following counterexamples.

\underline{{\it Counterexample 1}}

Assuming $B(t)\ne 0$ and using the new time variable $d\tau =\frac{dt}{B(t)}$
we rewrite the metric (\ref{sx1.1}) as: 
\begin{equation}
ds^2=B^2(\tau )\left[ -d\tau ^2+\frac{A^2(\tau )}{B^2(\tau )}%
dx^2+dy^2+\Sigma ^2(y,k)dz^2\right] .  \label{sx1.11}
\end{equation}
Consider a spacetime for which the components of $G_{ab}$ satisfy the
relations: 
\begin{equation}
G_1=-c_1^2\tilde{\tau}^2\qquad G_2=\pm c_2^2  \label{sx1.12}
\end{equation}
with: 
\begin{equation}
\tilde{\tau}(\tau )=\int \left| G_0\right| ^{1/2}d\tau  \label{sx1.13}
\end{equation}
where $k=0$ and $sign(G_0)>0$ (this case corresponds to the case $\alpha
_1\neq 0,\alpha _2=0$ of \cite{Camci-Sharif}).

It is easy to check that for this class of spacetimes (\ref{sx1.11}) we have
the following six {\em proper} MCs (dim${\cal C=}10$): 
\begin{equation}
{\bf X}_1=\cosh c_1x\partial _{\tilde{\tau}}-\frac 1{c_1\tilde{\tau}}\sinh
c_1x\partial _x  \label{sx1.13a}
\end{equation}
\begin{equation}
{\bf X}_2=\sinh c_1x\partial _{\tilde{\tau}}-\frac 1{c_1\tilde{\tau}}\cosh
c_1x\partial _x  \label{sx1.13b}
\end{equation}

\begin{equation}
{\bf X}_{2\mu +\nu }=-c_2f_{(\mu )}f_{(\nu )}^{\prime }\partial _{\tilde{\tau%
}}+\frac{c_2f_{(\mu )}\left[ f_{(\nu )}^{\prime }\right] _{,x}}{c_1^2\tilde{%
\tau}}\partial _x-\frac{\tilde{\tau}f_{(\nu )}^{\prime }\left[ f_{(\mu
)}\right] _{,y}}{c_2}\partial _y-\frac{\tilde{\tau}f_{(\nu )}^{\prime
}\left[ f_{(\mu )}\right] _{,z}}{y^2c_2}\partial _z  \label{sx1.14}
\end{equation}
where: 
\begin{equation}
f_{(\mu )}=\left( y\cos z,y\sin z\right)  \label{sx1.15}
\end{equation}
\begin{equation}
f_{(\nu )}^{\prime }=-\left( \cosh c_1x,\sinh c_1x\right)  \label{sx1.16}
\end{equation}
and the non tensorial indices $\mu =1,2$ and $\nu =1,2$ count vector fields.

These proper MCs are not given in \cite{Camci-Sharif}. Furthermore for $k=0$
the MCs they found (e.g. the vector fields ${\bf \xi }_9,{\bf \xi }_{10}$ in
equation (53) of \cite{Camci-Sharif}) are equal to zero, because when $%
k=0\Leftrightarrow T_2=$constant therefore $\dot{T}_2=0$.

\underline{{\it Counterexample 2}}

Consider the spherically/hyperbolic symmetric spacetime ($k=\pm 1$) in which
the metric functions are given by: 
\begin{equation}
A(\tau )=\frac{D_1}{D_2}\qquad B(\tau )=B_1\sinh ^2\frac{c_2\tau }2,B_1\sin
^2\frac{c_2\tau }2,B_1\cosh ^2\frac{c_2\tau }2  \label{sx1.17}
\end{equation}
where $D_1,D_2,B_1$ are constants of integration.

For these spacetimes the components of $G_{ab}$ satisfy the relations: 
\begin{equation}
G_1=-c_1^2c_2^2\qquad G_2=\pm c_2^2  \label{sx1.18}
\end{equation}
where $c_1^2=\pm \frac{D_1^2}{D_2^2}\frac{c_2^2\pm 1}{c_2^2}$ and the signs
depend on the forms of the metric function $B(\tau )$ and $k$.

Spacetimes (\ref{sx1.17}) are special cases of the class of metrics
satisfying (\ref{sx1.18}). They correspond to the case $\alpha _1=0,\alpha
_2=0$ of \cite{Camci-Sharif} for which the authors state that there do not
exist proper MCs. However it is easy to check that the following two vectors
are {\em proper} MCs (hence dim${\cal C=}6$): 
\begin{equation}
{\bf X}_1=\left| G_0\right| ^{-1/2}\partial _\tau \qquad {\bf X}_2=\left|
G_0\right| ^{-1/2}c_1c_2x\partial _\tau +\frac{\tilde{\tau}(\tau )}{c_1c_2}%
\partial _x  \label{sx1.19}
\end{equation}
where, as previously: 
\begin{equation}
\tilde{\tau}(\tau )=\int \left| G_0\right| ^{1/2}d\tau .  \label{sx1.20}
\end{equation}
We conclude this note by giving in Tables 1, 2,3,4 the {\em complete} Lie
Algebra of proper MCs for the metrics (\ref{sx1.1}). The results are given
in terms of the coordinate $\tilde{\tau}$ (essentially $G_0$) and some
integration constants. In the Tables the first column enumerates the various
cases, the second column gives the constant curvature of the spatial
2-space, the third and fourth columns give the corresponding forms of $%
G_1,G_2$, the fifth column gives the dimension of the Lie Algebra of MCs
(including the Lie algebra of the four Killing Vectors) and finally the
sixth column gives the form of the Collineation vectors. \vspace{0.3cm}

\begin{center}
{\bf {\small Table 1}.}{\small \ Matter Collineations admitted by the
metrics (\ref{sx1.11}). The sign of }$G_1${\small \ is such that sign}$%
(G_0\cdot G_1)<0${\small .} {\small \vspace{0.05cm}}

\begin{tabular}{|l|l|l|l|l|l|}
\hline
{\bf Class} & ${\bf k}$ & $G_1$ & $G_2$ & $\dim {\cal C}$ & ${\bf X}$ \\ 
\hline
$A_1$ & $0$ & $\pm c^2e^{-2\tilde{\tau}/\alpha _1c}$ & $\pm c^2e^{-2\tilde{%
\tau}/c}$ & {\bf 5} & $\alpha _1c\partial _{\tilde{\tau}}+x\partial
_x+\alpha _1y\partial _y$ \\ \hline
$A_2$ & $\pm 1$ & $\pm c_1^2c_2^2$ & $\pm c_2^2$ & {\bf 6} & $
\begin{array}{c}
\partial _{\tilde{\tau}} \\ 
c_1c_2x\partial _{\tilde{\tau}}+\frac{\tilde{\tau}}{c_1c_2}\partial _x
\end{array}
$ \\ \hline
$A_3$ & $0,\pm 1$ & $\pm c_1^2e^{\frac{2\tilde{\tau}}{ac_2}}$ & $\pm c_2^2$
& {\bf 6} & $
\begin{array}{c}
-ac_2\partial _{\tilde{\tau}}+x\partial _x \\ 
2ac_2x\partial _{\tilde{\tau}}-\left( x^2+\frac{a^2c_2^2}{c_1^2}e^{-\frac{2%
\tilde{\tau}}{ac_2}}\right) \partial _x
\end{array}
$ \\ \hline
$A_4$ & $0,\pm 1$ & $\pm c^2\cosh ^2\frac{\tilde{\tau}}{ac}$ & $\pm c^2$ & 
{\bf 6} & $
\begin{array}{c}
c\sin \frac xa\partial _{\tilde{\tau}}+\tanh \frac{\tilde{\tau}}{ac}\cos
\frac xa\partial _x \\ 
c\cos \frac xa\partial _{\tilde{\tau}}-\tanh \frac{\tilde{\tau}}{ac}\sin
\frac xa\partial _x
\end{array}
$ \\ \hline
$A_5$ & $0,\pm 1$ & $\pm c^2\sinh ^2\frac{\tilde{\tau}}{ac}$ & $\pm c^2$ & 
{\bf 6} & $
\begin{array}{c}
c\sinh \frac xa\partial _{\tilde{\tau}}-\coth \frac{\tilde{\tau}}{ac}\cosh
\frac xa\partial _x \\ 
c\cosh \frac xa\partial _{\tilde{\tau}}-\coth \frac{\tilde{\tau}}{ac}\sinh
\frac xa\partial _x
\end{array}
$ \\ \hline
$A_6$ & $0,\pm 1$ & $\pm c^2\cos ^2\frac{\tilde{\tau}}{ac}$ & $\pm c^2$ & 
{\bf 6} & $
\begin{array}{c}
c\sinh \frac xa\partial _{\tilde{\tau}}+\tan \frac{\tilde{\tau}}{ac}\cosh
\frac xa\partial _x \\ 
c\cosh \frac xa\partial _{\tilde{\tau}}+\tan \frac{\tilde{\tau}}{ac}\sinh
\frac xa\partial _x
\end{array}
$ \\ \hline
$A_7$ & $\pm 1$ & $\pm \tilde{\tau}^2$ & $\pm c^2$ & {\bf 6} & $
\begin{array}{c}
\cosh x\partial _{\tilde{\tau}}-\tilde{\tau}^{-1}\sinh x\partial _x \\ 
\sinh x\partial _{\tilde{\tau}}-\tilde{\tau}^{-1}\cosh x\partial _x
\end{array}
$ \\ \hline
\end{tabular}
\newpage

{\bf {\small Table 2}.}{\small \ Matter Collineations admitted by the
metrics (\ref{sx1.11}). The sign of }$G_1${\small \ is such that sign}$%
(G_0\cdot G_1)<0${\small .}

\begin{tabular}{|l|l|l|l|l|l|}
\hline
{\bf Class} & ${\bf k}$ & $G_1$ & $G_2$ & $\dim {\cal C}$ & ${\bf X}$ \\ 
\hline
$B_1$ & $1$ & $\pm c_1^2c^2$ & $\pm c^2\cosh ^2\frac{\tilde{\tau}}c$ & {\bf 7%
} & $
\begin{array}{c}
{\bf X}_{\mu +\nu +3}=-f_{(\mu )}\left[ f_{(\nu )}^{\prime }\right] _{,%
\tilde{\tau}}\left( c\cosh \frac{\tilde{\tau}}c\right) ^2\partial _{\tilde{%
\tau}}+ \\ 
+\frac{f_{(\mu )}\left[ f_{(\nu )}^{\prime }\right] _{,x}}{c_1^2\tanh ^2%
\frac{\tilde{\tau}}c}\partial _x-f_{(\nu )}^{\prime }\left[ f_{(\mu
)}\right] _{,y}\partial _y-\frac{f_{(\nu )}^{\prime }\left[ f_{(\mu
)}\right] _{,z}}{\sin ^2y}\partial _z
\end{array}
$ \\ \hline
$B_2$ & $-1$ & $\pm c_1^2c^2$ & $\pm c^2\sinh ^2\frac{\tilde{\tau}}c$ & {\bf %
7} & $
\begin{array}{c}
{\bf X}_{\mu +\nu +3}=-f_{(\mu )}\left[ f_{(\nu )}^{\prime }\right] _{,%
\tilde{\tau}}\left( c\sinh \frac{\tilde{\tau}}c\right) ^2\partial _{\tilde{%
\tau}}+ \\ 
+\frac{f_{(\mu )}\left[ f_{(\nu )}^{\prime }\right] _{,x}}{c_1^2\coth ^2%
\frac{\tilde{\tau}}c}\partial _x-f_{(\nu )}^{\prime }\left[ f_{(\mu
)}\right] _{,y}\partial _y-\frac{f_{(\nu )}^{\prime }\left[ f_{(\mu
)}\right] _{,z}}{\sinh ^2y}\partial _z
\end{array}
$ \\ \hline
$B_3$ & $-1$ & $\pm c_1^2c^2$ & $\pm c^2\sin ^2\frac{\tilde{\tau}}c$ & {\bf 7%
} & $
\begin{array}{c}
{\bf X}_{\mu +\nu +3}=-f_{(\mu )}\cdot \left[ f_{(\nu )}^{\prime }\right] _{,%
\tilde{\tau}}\left( c\sin \frac{\tilde{\tau}}c\right) ^2\partial _{\tilde{%
\tau}}+ \\ 
+\frac{f_{(\mu )}\left[ f_{(\nu )}^{\prime }\right] _{,x}}{c_1^2\cot ^2\frac{%
\tilde{\tau}}c}\partial _x-f_{(\nu )}^{\prime }\left[ f_{(\mu )}\right]
_{,y}\partial _y-\frac{f_{(\nu )}^{\prime }\left[ f_{(\mu )}\right] _{,z}}{%
\sinh ^2y}\partial _z
\end{array}
$ \\ \hline
$B_4$ & $1$ & $\pm c_1^2c^2\sinh ^2\frac{\tilde{\tau}}c$ & $\pm c^2\cosh ^2%
\frac{\tilde{\tau}}c$ & {\bf 10} & $
\begin{array}{c}
{\bf X}_{2(\mu +1)+\nu }=-f_{(\mu )}\left[ f_{(\nu )}^{\prime }\right] _{,%
\tilde{\tau}}\left( c\cosh \frac{\tilde{\tau}}c\right) ^2\partial _{\tilde{%
\tau}}+ \\ 
+\frac{f_{(\mu )}\left[ f_{(\nu )}^{\prime }\right] _{,x}}{c_1^2\tanh ^2%
\frac{\tilde{\tau}}c}\partial _x-f_{(\nu )}^{\prime }\left[ f_{(\mu
)}\right] _{,y}\partial _y-\frac{f_{(\nu )}^{\prime }\left[ f_{(\mu
)}\right] _{,z}}{\sin ^2y}\partial _z
\end{array}
$ \\ \hline
$B_5$ & $-1$ & $\pm c_1^2c^2\cosh ^2\frac{\tilde{\tau}}c$ & $\pm c^2\sinh ^2%
\frac{\tilde{\tau}}c$ & {\bf 10} & $
\begin{array}{c}
{\bf X}_{2(\mu +1)+\nu }=-f_{(\mu )}\left[ f_{(\nu )}^{\prime }\right] _{,%
\tilde{\tau}}\left( c\sinh \frac{\tilde{\tau}}c\right) ^2\partial _{\tilde{%
\tau}}+ \\ 
+\frac{f_{(\mu )}\left[ f_{(\nu )}^{\prime }\right] _{,x}}{c_1^2\coth ^2%
\frac{\tilde{\tau}}c}\partial _x-f_{(\nu )}^{\prime }\left[ f_{(\mu
)}\right] _{,y}\partial _y-\frac{f_{(\nu )}^{\prime }\left[ f_{(\mu
)}\right] _{,z}}{\sinh ^2y}\partial _z
\end{array}
$ \\ \hline
$B_6$ & $-1$ & $\pm c_1^2c^2\cos ^2\frac{\tilde{\tau}}c$ & $\pm c^2\sin ^2%
\frac{\tilde{\tau}}c$ & {\bf 10} & $
\begin{array}{c}
{\bf X}_{2(\mu +1)+\nu }=-f_{(\mu )}\left[ f_{(\nu )}^{\prime }\right] _{,%
\tilde{\tau}}\left( c\sin \frac{\tilde{\tau}}c\right) ^2\partial _{\tilde{%
\tau}}+ \\ 
+\frac{f_{(\mu )}\left[ f_{(\nu )}^{\prime }\right] _{,x}}{c_1^2\cot ^2\frac{%
\tilde{\tau}}c}\partial _x-f_{(\nu )}^{\prime }\left[ f_{(\mu )}\right]
_{,y}\partial _y-\frac{f_{(\nu )}^{\prime }\left[ f_{(\mu )}\right] _{,z}}{%
\sinh ^2y}\partial _z
\end{array}
$ \\ \hline
$B_7$ & $0$ & $\pm c_1^2$ & $\pm c_2^2$ & {\bf 10} & $
\begin{array}{c}
{\bf X}_{2(\mu +1)+\nu }=-c_2f_{(\mu )}\left[ f_{(\nu )}^{\prime }\right] _{,%
\tilde{\tau}}\partial _{\tilde{\tau}}+\frac{c_2f_{(\mu )}\left[ f_{(\nu
)}^{\prime }\right] _{,x}}{c_1^2}\partial _x- \\ 
-\frac{f_{(\nu )}^{\prime }\left[ f_{(\mu )}\right] _{,y}}{c_2}\partial _y-%
\frac{f_{(\nu )}^{\prime }\left[ f_{(\mu )}\right] _{,z}}{y^2c_2}\partial _z
\\ 
{\bf X}_9=\partial _{\tilde{\tau}} \\ 
{\bf X}_{10}=c_1x\partial _{\tilde{\tau}}+\frac{\tilde{\tau}}{c_1}\partial _x
\end{array}
$ \\ \hline
$B_8$ & $0$ & $\pm c_1^2\tilde{\tau}^2$ & $\pm c_2^2$ & {\bf 10} & $
\begin{array}{c}
{\bf X}_{2(\mu +1)+\nu }=-c_2f_{(\mu )}f_{(\nu )}^{\prime }\partial _{\tilde{%
\tau}}+\frac{c_2f_{(\mu )}\left[ f_{(\nu )}^{\prime }\right] _{,x}}{c_1^2%
\tilde{\tau}}\partial _x- \\ 
-\frac{\tilde{\tau}f_{(\nu )}^{\prime }\left[ f_{(\mu )}\right] _{,y}}{c_2}%
\partial _y-\frac{\tilde{\tau}f_{(\nu )}^{\prime }\left[ f_{(\mu )}\right]
_{,z}}{y^2c_2}\partial _z \\ 
{\bf X}_9=\cosh c_1x\partial _{\tilde{\tau}}-\frac 1{c_1\tilde{\tau}}\sinh
c_1x\partial _x \\ 
{\bf X}_{10}=\sinh c_1x\partial _{\tilde{\tau}}-\frac 1{c_1\tilde{\tau}%
}\cosh c_1x\partial _x
\end{array}
$ \\ \hline
\end{tabular}
\newpage

{\bf {\small Table 3}. }{\small Matter Collineations admitted by the metrics
(\ref{sx1.11}). The sign of }$G_1${\small \ is such that sign}$(G_0\cdot
G_1)>0${\small . \vspace{0.1cm}}

\begin{tabular}{|l|l|l|l|l|l|}
\hline
{\bf Class} & ${\bf k}$ & $G_1$ & $G_2$ & $\dim {\cal C}$ & ${\bf X}$ \\ 
\hline
$A_1$ & $0$ & $\pm c^2e^{-2\tilde{\tau}/\alpha _1c}$ & $\pm c^2e^{-2\tilde{%
\tau}/c}$ & {\bf 5} & $\alpha _1c\partial _{\tilde{\tau}}+x\partial
_x+\alpha _1y\partial _y$ \\ \hline
$A_2$ & $\pm 1$ & $\pm c_1^2c_2^2$ & $\pm c_2^2$ & {\bf 6} & $
\begin{array}{c}
\partial _{\tilde{\tau}} \\ 
c_1c_2x\partial _{\tilde{\tau}}-\frac{\tilde{\tau}}{c_1c_2}\partial _x
\end{array}
$ \\ \hline
$A_3$ & $0,\pm 1$ & $\pm c_1^2e^{\frac{2\tilde{\tau}}{ac_2}}$ & $\pm c_2^2$
& {\bf 6} & $
\begin{array}{c}
-ac_2\partial _{\tilde{\tau}}+x\partial _x \\ 
2ac_2x\partial _{\tilde{\tau}}-\left( x^2-\frac{a^2c_2^2}{c_1^2}e^{-\frac{2%
\tilde{\tau}}{ac_2}}\right) \partial _x
\end{array}
$ \\ \hline
$A_4$ & $0,\pm 1$ & $\pm c^2\cos ^2\frac{\tilde{\tau}}{ac}$ & $\pm c^2$ & 
{\bf 6} & $
\begin{array}{c}
c\sin \frac xa\partial _{\tilde{\tau}}-\tan \frac{\tilde{\tau}}{ac}\cos
\frac xa\partial _x \\ 
c\cos \frac xa\partial _{\tilde{\tau}}+\tan \frac{\tilde{\tau}}{ac}\sin
\frac xa\partial _x
\end{array}
$ \\ \hline
$A_5$ & $0,\pm 1$ & $\pm c^2\sinh ^2\frac{\tilde{\tau}}{ac}$ & $\pm c^2$ & 
{\bf 6} & $
\begin{array}{c}
c\sin \frac xa\partial _{\tilde{\tau}}+\coth \frac{\tilde{\tau}}{ac}\cos
\frac xa\partial _x \\ 
c\cos \frac xa\partial _{\tilde{\tau}}-\coth \frac{\tilde{\tau}}{ac}\sin
\frac xa\partial _x
\end{array}
$ \\ \hline
$A_6$ & $0,\pm 1$ & $\pm c^2\cosh ^2\frac{\tilde{\tau}}{ac}$ & $\pm c^2$ & 
{\bf 6} & $
\begin{array}{c}
c\sinh \frac xa\partial _{\tilde{\tau}}-\tanh \frac{\tilde{\tau}}{ac}\cosh
\frac xa\partial _x \\ 
c\cosh \frac xa\partial _{\tilde{\tau}}-\tanh \frac{\tilde{\tau}}{ac}\sinh
\frac xa\partial _x
\end{array}
$ \\ \hline
$A_7$ & $\pm 1$ & $\pm \tilde{\tau}^2$ & $\pm c^2$ & {\bf 6} & $
\begin{array}{c}
\cos x\partial _{\tilde{\tau}}-\tilde{\tau}^{-1}\sin x\partial _x \\ 
\sin x\partial _{\tilde{\tau}}+\tilde{\tau}^{-1}\cos x\partial _x
\end{array}
$ \\ \hline
\end{tabular}
\newpage

{\bf {\small Table 4}. }{\small Matter Collineations admitted by the metrics
(\ref{sx1.11}). The sign of }$G_1${\small \ is such that sign}$(G_0\cdot
G_1)>0${\small . }

\begin{tabular}{|l|l|l|l|l|l|}
\hline
{\bf Class} & ${\bf k}$ & $G_1$ & $G_2$ & $\dim {\cal C}$ & ${\bf X}$ \\ 
\hline
$B_1$ & $1$ & $\pm c_1^2c^2$ & $\pm c^2\cos ^2\frac{\tilde{\tau}}c$ & {\bf 7}
& $
\begin{array}{c}
{\bf X}_{\mu +\nu +3}=f_{(\mu )}\left[ f_{(\nu )}^{\prime }\right] _{,\tilde{%
\tau}}\left( c\cos \frac{\tilde{\tau}}c\right) ^2\partial _{\tilde{\tau}}+
\\ 
+\frac{f_{(\mu )}\left[ f_{(\nu )}^{\prime }\right] _{,x}}{c_1^2\tan ^2\frac{%
\tilde{\tau}}c}\partial _x-f_{(\nu )}^{\prime }\left[ f_{(\mu )}\right]
_{,y}\partial _y-\frac{f_{(\nu )}^{\prime }\left[ f_{(\mu )}\right] _{,z}}{%
\sin ^2y}\partial _z
\end{array}
$ \\ \hline
$B_2$ & $-1$ & $\pm c_1^2c^2$ & $\pm c^2\cosh ^2\frac{\tilde{\tau}}c$ & {\bf %
7} & $
\begin{array}{c}
{\bf X}_{\mu +\nu +3}=f_{(\mu )}\left[ f_{(\nu )}^{\prime }\right] _{,\tilde{%
\tau}}\left( c\cosh \frac{\tilde{\tau}}c\right) ^2\partial _{\tilde{\tau}}+
\\ 
+\frac{f_{(\mu )}\left[ f_{(\nu )}^{\prime }\right] _{,x}}{c_1^2\tanh ^2%
\frac{\tilde{\tau}}c}\partial _x-f_{(\nu )}^{\prime }\left[ f_{(\mu
)}\right] _{,y}\partial _y-\frac{f_{(\nu )}^{\prime }\left[ f_{(\mu
)}\right] _{,z}}{\sinh ^2y}\partial _z
\end{array}
$ \\ \hline
$B_3$ & $1$ & $\pm c_1^2c^2$ & $\pm c^2\sinh ^2\frac{\tilde{\tau}}c$ & {\bf 7%
} & $
\begin{array}{c}
{\bf X}_{\mu +\nu +3}=f_{(\mu )}\left[ f_{(\nu )}^{\prime }\right] _{,\tilde{%
\tau}}\left( c\sinh \frac{\tilde{\tau}}c\right) ^2\partial _{\tilde{\tau}}+
\\ 
+\frac{f_{(\mu )}\left[ f_{(\nu )}^{\prime }\right] _{,x}}{c_1^2\coth ^2%
\frac{\tilde{\tau}}c}\partial _x-f_{(\nu )}^{\prime }\left[ f_{(\mu
)}\right] _{,y}\partial _y-\frac{f_{(\nu )}^{\prime }\left[ f_{(\mu
)}\right] _{,z}}{\sin ^2y}\partial _z
\end{array}
$ \\ \hline
$B_4$ & $1$ & $\pm c_1^2c^2\sin ^2\frac{\tilde{\tau}}c$ & $\pm c^2\cos ^2%
\frac{\tilde{\tau}}c$ & {\bf 10} & $
\begin{array}{c}
{\bf X}_{2(\mu +1)+\nu }=f_{(\mu )}\left[ f_{(\nu )}^{\prime }\right] _{,%
\tilde{\tau}}\left( c\cos \frac{\tilde{\tau}}c\right) ^2\partial _{\tilde{%
\tau}}+ \\ 
+\frac{f_{(\mu )}\left[ f_{(\nu )}^{\prime }\right] _{,x}}{c_1^2\tan ^2\frac{%
\tilde{\tau}}c}\partial _x-f_{(\nu )}^{\prime }\left[ f_{(\mu )}\right]
_{,y}\partial _y-\frac{f_{(\nu )}^{\prime }\left[ f_{(\mu )}\right] _{,z}}{%
\sin ^2y}\partial _z
\end{array}
$ \\ \hline
$B_5$ & $-1$ & $\pm c_1^2c^2\sinh ^2\frac{\tilde{\tau}}c$ & $\pm c^2\cosh ^2%
\frac{\tilde{\tau}}c$ & {\bf 10} & $
\begin{array}{c}
{\bf X}_{2(\mu +1)+\nu }=f_{(\mu )}\left[ f_{(\nu )}^{\prime }\right] _{,%
\tilde{\tau}}\left( c\cosh \frac{\tilde{\tau}}c\right) ^2\partial _{\tilde{%
\tau}}+ \\ 
+\frac{f_{(\mu )}\left[ f_{(\nu )}^{\prime }\right] _{,x}}{c_1^2\tanh ^2%
\frac{\tilde{\tau}}c}\partial _x-f_{(\nu )}^{\prime }\left[ f_{(\mu
)}\right] _{,y}\partial _y-\frac{f_{(\nu )}^{\prime }\left[ f_{(\mu
)}\right] _{,z}}{\sinh ^2y}\partial _z
\end{array}
$ \\ \hline
$B_6$ & $1$ & $\pm c_1^2c^2\cosh ^2\frac{\tilde{\tau}}c$ & $\pm c^2\sinh ^2%
\frac{\tilde{\tau}}c$ & {\bf 10} & $
\begin{array}{c}
{\bf X}_{2(\mu +1)+\nu }=f_{(\mu )}\left[ f_{(\nu )}^{\prime }\right] _{,%
\tilde{\tau}}\left( c\sinh \frac{\tilde{\tau}}c\right) ^2\partial _{\tilde{%
\tau}}+ \\ 
+\frac{f_{(\mu )}\left[ f_{(\nu )}^{\prime }\right] _{,x}}{c_1^2\coth ^2%
\frac{\tilde{\tau}}c}\partial _x-f_{(\nu )}^{\prime }\left[ f_{(\mu
)}\right] _{,y}\partial _y-\frac{f_{(\nu )}^{\prime }\left[ f_{(\mu
)}\right] _{,z}}{\sin ^2y}\partial _z
\end{array}
$ \\ \hline
$B_7$ & $0$ & $\pm c_1^2$ & $\pm c_2^2$ & {\bf 10} & $
\begin{array}{c}
{\bf X}_{2(\mu +1)+\nu }=c_2f_{(\mu )}\left[ f_{(\nu )}^{\prime }\right] _{,%
\tilde{\tau}}\partial _{\tilde{\tau}}+\frac{c_2f_{(\mu )}\left[ f_{(\nu
)}^{\prime }\right] _{,x}}{c_1^2}\partial _x- \\ 
-\frac{f_{(\nu )}^{\prime }\left[ f_{(\mu )}\right] _{,y}}{c_2}\partial _y-%
\frac{f_{(\nu )}^{\prime }\left[ f_{(\mu )}\right] _{,z}}{y^2c_2}\partial _z
\\ 
{\bf X}_9=\partial _{\tilde{\tau}} \\ 
{\bf X}_{10}=c_1x\partial _{\tilde{\tau}}-\frac{\tilde{\tau}}{c_1}\partial _x
\end{array}
$ \\ \hline
$B_8$ & $0$ & $\pm c_1^2\tilde{\tau}^2$ & $\pm c_2^2$ & {\bf 10} & $
\begin{array}{c}
{\bf X}_{2(\mu +1)+\nu }=c_2f_{(\mu )}f_{(\nu )}^{\prime }\partial _{\tilde{%
\tau}}+\frac{c_2f_{(\mu )}\left[ f_{(\nu )}^{\prime }\right] _{,x}}{c_1^2%
\tilde{\tau}}\partial _x- \\ 
-\frac{\tilde{\tau}f_{(\nu )}^{\prime }\left[ f_{(\mu )}\right] _{,y}}{c_2}%
\partial _y-\frac{\tilde{\tau}f_{(\nu )}^{\prime }\left[ f_{(\mu )}\right]
_{,z}}{y^2c_2}\partial _z \\ 
{\bf X}_9=\cos c_1x\partial _{\tilde{\tau}}-\frac 1{c_1\tilde{\tau}}\sin
c_1x\partial _x \\ 
{\bf X}_{10}=\sin c_1x\partial _{\tilde{\tau}}+\frac 1{c_1\tilde{\tau}}\cos
c_1x\partial _x
\end{array}
$ \\ \hline
\end{tabular}
\newpage {\bf {\small Table 5}. }{\small Explanations for the quantities }$%
f_{(\mu )},f_{(\nu )}^{\prime }${\small \ appearing in Table 2.}$\mu ,\nu
=1,2,3$. \vspace{0.1cm}

\begin{tabular}{|l|l|l|l|}
\hline
{\bf Class} & ${\bf k}$ & $f_{(\nu )}^{\prime }$ & $f_{(\mu )}$ \\ \hline
$B_1$ & $1$ & $\left( -\tanh \frac{\tilde{\tau}}c,0,0\right) $ & $\left(
-\cos y,\sin y\cos z,\sin y\sin z\right) $ \\ \hline
$B_2$ & $-1$ & $\left( \coth \frac{\tilde{\tau}}c,0,0\right) $ & $\left(
\cosh y,\sinh y\cos z,\sinh y\sin z\right) $ \\ \hline
$B_3$ & $-1$ & $\left( -\cot \frac{\tilde{\tau}}c,0,0\right) $ & $\left(
\cosh y,\sinh y\cos z,\sinh y\sin z\right) $ \\ \hline
$B_4$ & $1$ & $\left( \tanh \frac{\tilde{\tau}}c\cosh c_1x,\tanh \frac{%
\tilde{\tau}}c\sinh c_1x,0\right) $ & $\left( -\cos y,\sin y\cos z,\sin
y\sin z\right) $ \\ \hline
$B_5$ & $-1$ & $\left( -\coth \frac{\tilde{\tau}}c\cos c_1x,-\coth \frac{%
\tilde{\tau}}c\sin c_1x,0\right) $ & $\left( \cosh y,\sinh y\cos z,\sinh
y\sin z\right) $ \\ \hline
$B_6$ & $-1$ & $\left( -\cot \frac{\tilde{\tau}}c\cosh c_1x,-\cot \frac{%
\tilde{\tau}}c\sinh c_1x,0\right) $ & $\left( \cosh y,\sinh y\cos z,\sinh
y\sin z\right) $ \\ \hline
$B_7$ & $0$ & $-\left( \tilde{\tau},c_1x,0\right) $ & $\left( y\cos z,y\sin
z,0\right) $ \\ \hline
$B_8$ & $0$ & $-\left( \cosh c_1x,\sinh c_1x,0\right) $ & $\left( y\cos
z,y\sin z,0\right) $ \\ \hline
\end{tabular}
\end{center}

\vspace{0.2cm}

\begin{center}
{\bf {\small Table 6}. }{\small Explanations for the quantities }$f_{(\mu
)},f_{(\nu )}^{\prime }${\small \ appearing in Table 4. }$\mu ,\nu =1,2,3$. 
\vspace{0.1cm}

\begin{tabular}{|l|l|l|l|}
\hline
{\bf Class} & ${\bf k}$ & $f_{(\nu )}^{\prime }$ & $f_{(\mu )}$ \\ \hline
$B_1$ & $1$ & $\left( -\tan \frac{\tilde{\tau}}c,0,0\right) $ & $\left(
-\cos y,\sin y\cos z,\sin y\sin z\right) $ \\ \hline
$B_2$ & $-1$ & $\left( \tanh \frac{\tilde{\tau}}c,0,0\right) $ & $\left(
\cosh y,\sinh y\cos z,\sinh y\sin z\right) $ \\ \hline
$B_3$ & $1$ & $\left( -\coth \frac{\tilde{\tau}}c,0,0\right) $ & $\left(
\cos y,\sin y\cos z,\sin y\sin z\right) $ \\ \hline
$B_4$ & $1$ & $\left( \tan \frac{\tilde{\tau}}c\cos c_1x,\tan \frac{\tilde{%
\tau}}c\sin c_1x,0\right) $ & $\left( -\cos y,\sin y\cos z,\sin y\sin
z\right) $ \\ \hline
$B_5$ & $-1$ & $\left( -\tanh \frac{\tilde{\tau}}c\cos c_1x,-\tanh \frac{%
\tilde{\tau}}c\sin c_1x,0\right) $ & $\left( \cosh y,\sinh y\cos z,\sinh
y\sin z\right) $ \\ \hline
$B_6$ & $1$ & $\left( -\coth \frac{\tilde{\tau}}c\cosh c_1x,-\coth \frac{%
\tilde{\tau}}c\sinh c_1x,0\right) $ & $\left( \cos y,\sin y\cos z,\sin y\sin
z\right) $ \\ \hline
$B_7$ & $0$ & $-\left( \tilde{\tau},c_1x,0\right) $ & $\left( y\cos z,y\sin
z,0\right) $ \\ \hline
$B_8$ & $0$ & $-\left( \cos c_1x,\sin c_1x,0\right) $ & $\left( y\cos
z,y\sin z,0\right) $ \\ \hline
\end{tabular}
\end{center}

\vspace{0.5cm}

A systematic and complete study of MCs in hypersurface LRS spacetimes (which
includes the present case as a special case) will be discussed in a
forthcoming work.

{\bf Note added:}

In order the spacetimes (1) to admit a MC, the spatial components $G_1,G_2$
of the Einstein tensor must satisfy a first order differential equation
whose solution gives $G_1,G_2$ and, consequently, the collineation vectors.
A detailed presentation of these differential equations for {\em all}
hypersurface homogeneous LRS will be given in \cite{apostol-tsamp9}.

Each algebraic constraint (third and fourth column of Tables 1,2,3,4) leads
to a system of two differential equations among the metric functions $%
A(t),B(t),$ which, in general, is difficult to solve explicitly. In the
counterexamples for simplicity and in order to present spacetimes which are
not immediately ruled out as unphysical (in fact it can be shown that the
spacetimes (19) satisfy all the energy conditions) we use the most simple
case where $G_1,G_2$ are constants. In this case a class of solutions of the
system of differential equations are the metric functions (19) with $%
D_1,D_2,B_1$ being constants of integration.

\end{document}